\documentclass[aip,pop,reprint]{revtex4-1}
\usepackage{dcolumn}
\usepackage{amsmath}
\usepackage{bm}
\usepackage{hyperref}
\usepackage{natbib}
\usepackage{amsfonts}
\usepackage{booktabs}
\usepackage{siunitx} 
\usepackage{relsize}
\usepackage{mathtools}
\usepackage[colorinlistoftodos]{todonotes}
\usepackage{physics}
\usepackage{graphicx}
\usepackage{array}
\usepackage{float}
\usepackage{svg}
\usepackage{tikz}
\usetikzlibrary{tikzmark}
\usepackage{empheq}
\usepackage{enumitem}
\makeatletter
\def\namedlabel#1#2{\begingroup
    #2%
    \def\@currentlabel{#2}%
    \phantomsection\label{#1}\endgroup
}

\makeatother
\usepackage{tcolorbox}
\usepackage{color}
\usepackage{amssymb}

\begin{document}

\title{When should PIC simulations be applied to atmospheric pressure plasmas? Impact of correlation heating}

\author{M. D. Acciarri}
\altaffiliation[Corresponding author: ]{M. D. Acciarri}
\email{acciarri@umich.edu}
\address{Department of Nuclear Engineering and Radiological Sciences, University of Michigan, Ann Arbor, MI 48109, USA}
\author{C. Moore} 
\address{Sandia National Laboratories, Albuquerque, NM 87185, USA}
\author{L. P. Beving} 
\address{Sandia National Laboratories, Albuquerque, NM 87185, USA}
\author{S. D. Baalrud}
\address{Department of Nuclear Engineering and Radiological Sciences, University of Michigan, Ann Arbor, MI 48109, USA}

\date{\today}

\begin{abstract}

Molecular dynamics simulations are used to test when the particle-in-cell (PIC) method applies to atmospheric pressure plasmas. It is found that PIC applies only when the plasma density and macroparticle weight are sufficiently small because of two effects associated with correlation heating. The first is the physical effect of disorder-induced heating (DIH). This occurs if the plasma density is large enough that a species (typically ions) is strongly correlated in the sense that the Coulomb coupling parameter exceeds one. In this situation, DIH causes ions to rapidly heat following ionization. PIC is not well suited to capture DIH because doing so requires using a macroparticle weight of one and a grid that well resolves the physical interparticle spacing. These criteria render PIC intractable for macroscale domains. The second effect is a numerical error due to Artificial Correlation Heating (ACH).  ACH is like DIH in that it is caused by the Coulomb repulsion between particles, but differs in that it is a numerical effect caused by a macroparticle weight larger than one. Like DIH, it is associated with strong correlations. However, here the macroparticle coupling strength is found to scale as $\Gamma w^{2/3}$, where $\Gamma$ is the physical coupling strength and $w$ is the macroparticle weight. So even if the physical coupling strength of a species is small, as is expected for electrons in atmospheric pressure plasmas, a sufficiently large macroparticle weight can cause the macroparticles to be strongly coupled and therefore heat due to ACH. Furthermore, it is shown that simulations in reduced dimensions exacerbate these issues.

\end{abstract}


\maketitle
%------------------------------------------------------

\section{Introduction}

Low-temperature plasmas at atmospheric pressure and beyond possess immense potential for a broad range of applications, including medicine \cite{app11114809}, agriculture \cite{misra_cold_2016}, the food industry \cite{https://doi.org/10.1002/ppap.201700085}, water purification \cite{Adamovich_2022}, CO$_2$ conversion \cite{10.3389/fenrg.2020.00111}, plasma-assisted ignition (PAI) \cite{STARIKOVSKIY201361,Starikovskaia_2014}, and plasma-assisted combustion (PAC) \cite{STARIKOVSKIY201361,Starikovskaia_2014}. Given the transformative impact of these applications, understanding and controlling the characteristics of these plasmas is of paramount importance. Accurate and efficient modeling of plasma discharges at atmospheric and higher pressures is critical for the advancement and optimization of plasma devices. Simulations, for instance, can aid in identifying optimal operating parameters like discharge voltage, power, frequency, electrode configuration, and gas mixture for diverse applications. Consequently, it is imperative to develop reliable and efficient computational tools for simulating the behavior of cold atmospheric pressure plasmas.

The particle-in-cell (PIC) method is a commonly used computational approach in plasma modeling \cite{Verboncoeur_2005,Birdsall,hockney_eastwood_1989} that is primarily characterized by its use of numerical macroparticles. These macroparticles serve as aggregates of physical particles, allowing the method to simulate macroscopic scales effectively. However, PIC requires additional numerical conditions to ensure validity. First, the Debye length must be adequately resolved to prevent artificial “PIC heating” \cite{Birdsall,Chacon_2020}. Second, a sufficiently large number of macroparticles per cell must be maintained to ensure a statistically representative model of the plasma. In addition, the CFL condition must be met to resolve the corresponding plasma frequency. Adherence to these guidelines enables the PIC method to make kinetic simulations tractable and to simulate entire plasma devices with a reasonable computational cost. The aim of this work is to test the applicability of these underlying assumptions in the context of atmospheric pressure plasmas using molecular dynamics (MD) simulations. Molecular dynamics is used as the benchmark because it directly solves Newton's equations for all interactions between particles, includes strong coupling effects, and is therefore a more first-principles solution than PIC.

It is found that PIC simulations are not well-suited to atmospheric pressure plasmas when the plasma density or macroparticle weight are too large because of effects associated with correlation heating. Here, we refer to ``PIC simulations'' in the traditional sense, not including the P3M method~\cite{MD_Frenkel}, which we consider to be a variant of MD in this work. Recent work \cite{Acciarri2022} has revealed that ions at atmospheric pressure often exist in a strongly correlated regime. This correlation strength is quantified by the ion-ion coupling parameter $\Gamma_{ii}>1$, defined as the ratio of the average potential energy of interacting ions to their average kinetic energy,
\begin{equation}
    \Gamma_{ii} = \frac{Z^2 e^2/a_{ii}}{4\pi \epsilon_{0} k_{B} T_{i}}.
    \label{eq:gamma}
\end{equation} 
Here, $Z$ is the ion charge state, $T_i$ is the ion temperature and $a_{ii}=(3/4\pi n_i)^{1/3}$ is the mean ion separation. The assumptions of the PIC method presume weak coupling between charged particles, ignoring interactions within a macroparticle and within a cell. Hence, pure PIC simulations effectively solve the Vlasov equation. In an attempt to incorporate collisions, PIC is often combined with a Boltzmann collision operator through the MCC or DSMC methods \cite{Birdsall,hockney_eastwood_1989}. It may be expected that creating a collision routine to incorporate strong correlation effects could enable the successful application of PIC in these scenarios. However, it is shown here that PIC faces inherent challenges when modeling strongly coupled plasmas that can not be remedied by the addition of a collision routine. These shortcomings stem from the requirement that PIC simulations at atmospheric pressure need to resolve very small spatial scales (here the Debye length and mean ion separation) while using macroparticles to remain practical.

In fact, a primary factor precluding the effective use of PIC in these scenarios is a phenomenon known as disorder-induced heating (DIH). This physical effect is the consequence of energy conservation, where the Coulomb potential energy of ions is converted to kinetic energy as they move to the lowest potential configuration following ionization. Previous MD simulations have depicted this heating effect in ions following ionization in atmospheric pressure plasmas \cite{Acciarri2022}. Here it is shown that to correctly capture DIH within PIC simulations, it is necessary to adopt a macroparticle weight ($w$) of one and to resolve the physical interparticle spacing. However, adhering to these conditions renders PIC prohibitively expensive for macroscopic$-$scale domains, as the computational load dramatically increases. Compounding the challenge, it is observed that at strong coupling, the average interparticle spacing between ions is larger than the Debye length $\lambda_\textrm{D}$. This implies that in adhering to the requisite of the Debye length resolution, the physical interparticle spacing also needs to be resolved. This leads to less than one macroparticle per cell, inducing PIC heating even if the Debye length is resolved. The combination of these conditions significantly exacerbates the difficulty of employing PIC for the simulation of strongly correlated, atmospheric pressure plasmas.

Beyond the limitations associated with strongly coupled plasmas, this study reveals a new numerical heating mechanism, termed “artificial correlation heating” (ACH), which can arise even in weakly coupled plasmas. This mechanism is analogous to the cause of DIH, as it stems from Coulomb repulsion, but distinguishes itself as a numerical effect associated with a macroparticle weight larger than one, $w>1$. Interestingly, like DIH, ACH is also associated with a conversion of potential to kinetic energy that arises at strong coupling. However, the coupling strength involved in this case is a macroparticle coupling strength
\begin{equation}
    \Gamma^w_{ss^\prime} = \Gamma_{ss^\prime} w^{2/3} ,
    \label{eq:gamma_w_intro}
    \end{equation}
where $\Gamma_{ss^\prime}$ is the physical coupling strength for interactions between species $s$ and $s^\prime$, derived in equation \ref{eq:gamma_w}.
Thus, even if the physical coupling strength is small, the macroparticle coupling strength can be large if $w\gg 1$.
Avoiding ACH requires maintaining numerous macroparticles per cell, a factor that further complicates the application of the PIC method. 

In the context of atmospheric pressure plasmas, ions may reach a strongly correlated regime and therefore be influenced by DIH, while electrons are expected to be weakly coupled due to their larger temperature. 
This may lead to the expectation that PIC applies to the electrons. 
Although that can be true, we find that ACH is an important consideration if a large macroparticle weight is applied to the electrons. We find that ACH imposes an upper limit on the applicability of PIC simulations in terms of the electron density and macroparticle weight, and a minimum on the electron temperature and grid spacing. Accordingly, a model for ACH has been developed that compares well with PIC simulations, offering a tool to anticipate when this new constraint influences PIC simulations.

The paper unfolds as follows. In Section \ref{sec:simulation_setup}, the simulation setups for both MD and PIC are detailed. Section \ref{sec:DIH} presents the results from an MD simulation and provides a comparative analysis with PIC simulations, focusing on the influence of the macroparticle weight and grid resolution in strongly coupled ion scenarios. The implications of strong ion-ion correlations on PIC heating, along with considerations for running PIC simulations on reduced spatial dimensions, are subsequently discussed. Finally, Section \ref{sec:ACH} introduces the concept of ACH, underscoring its significance as a new constraint on the applicability of the PIC method.

\section{Simulation Setup}\label{sec:simulation_setup}

Molecular dynamics simulations were used to test the applicability of the PIC method. In both cases, the simulation domain consisted of a cubic box with periodic boundary conditions. The size of the simulation domain was calculated such that for a given number of particles, the particle density met a desired value. The basic molecular dynamics and PIC setups are described in sections \ref{subsec:simulation_setup_MD} and \ref{subsec:simulation_setup_PIC} respectively.

\subsection{Molecular Dynamics}\label{subsec:simulation_setup_MD}

Molecular dynamics simulations were carried out using the open-source software LAMMPS \cite{LAMMPS}. Since electrons are much hotter than ions and are weakly coupled, they are treated as a background non-interacting species when modeling ion dynamics. Thus, they were not included in the simulations. Furthermore, since the objective of this work is to test the limitations of PIC simulations when ions are strongly coupled, no ion-neutral interactions were included in the MD simulations. That is, only a one-component plasma model~\cite{BausPR1980} was simulated, which is known to provide an accurate description of ions in the presence of weakly coupled electrons, such as in ultracold neutral plasmas~\cite{KILLIAN200777}. In order to study the evolution of a non-equilibrium discharge, a neutral Ar gas at room temperature and atmospheric pressure was simulated until equilibrium was reached. This stage of the simulation was run with a Nos\'{e}-Hoover thermostat (NVT ensemble) applied \cite{MD_Frenkel} and the Lennard Jones potential \cite{Acciarri2022}. Then, the entire set of particles was instantly ionized and a NVE (microcanonical) simulation was run where ion-ion interactions were modeled using the Coulomb potential
\begin{equation}
    \phi(r) =  \frac{Z^2 e^2}{4 \pi \epsilon_0} \frac{1}{r}
    \label{eq:coulomb}
\end{equation} where $Ze$ is the ion charge, $\epsilon_0$ is the vacuum permittivity and $r$ is the radial distance. The P3M method was used to include both short and long range contributions in ion-ion interactions.~\cite{MD_Frenkel} A distance of $r_c=10\;a_{ii}$ was chosen to separate short and long range parts of the Ewald summation. The timestep used was $10^{-3}\omega_{pi}$ where $\omega_{pi}$ is the ion plasma period. The number of particles was 5000. The simulated ion density was  $2.5 \times 10^{24}$~m$^{-3}$ which is equivalent to the ion density of an atmospheric pressure plasma with an ionization fraction of 10\%.
This value was chosen as an example because DIH is significant, and it is relevant to a number experiments \cite{Lo_2017,vanderHorst2012}. Furthermore, a lower ion density within the strongly coupled regime could have been used and the same analysis would still apply \cite{Acciarri2022}.

\subsection{Particle in Cell}\label{subsec:simulation_setup_PIC}

Here, an in house developed electrostatic PIC code was utilized. The code adheres to the conventional method of interpolating particle charges onto a structured uniform grid using shape functions. Subsequently, field equations are solved, and the resulting fields are interpolated back to the particle positions to integrate the equations of motion. This process is repeated throughout each iteration of the PIC simulation \cite{Birdsall}. 

For the Poisson equation solver, the charge density was computed using the ion density and a uniform background neutralizing electron density to ensure stability
\begin{equation}
    \rho(\textbf{r}) = e \left( Z n_i(\textbf{r}) - n_e \right),
    \label{eq:rho}
\end{equation} where $\rho$ is the charge density, $Z$ is the ion charge state, $n_i$ is the ion density and $n_e$ is the background electron density assumed to be constant and equal to $\int_V n_i(\textbf{r}) d^3 r / V$ where $V$ is the volume of the simulation domain. The Poisson equation for the electrostatic potential was then solved using the spectral method. That is, by applying the Fourier transform and solving the consequent algebraic equation for the electric potential in the k-space,
\begin{subequations}
\begin{eqnarray}
  \label{eq:Poisson}
    \nabla^2 \phi(\textbf{r}) &=& \frac{- \rho(\textbf{r})}{\epsilon_0} \\      
     \phi_{\textbf{k}} 
     &=& \frac{\rho_{\textbf{k}}}{\epsilon_0 k^2}
     \\
     F^{-1}\left(\phi_{\textbf{k}}\right) &=& \phi(\textbf{r})
\end{eqnarray}
\end{subequations}
where $\phi$ is the electrostatic potential, $\epsilon_0$ is the vacuum permittivity and $F^{-1}$ is the inverse Fourier transform. The electric field components were obtained by numerically differentiating the electric potential among each direction using finite central differences of order $O(2)$ and periodic boundary conditions. No external electric fields were included.

Interpolation of the particle charges to the grid and the electric field from the grid to the particle positions was performed using the “scatter” and “gather” operations
\begin{equation}
    \rho_{i,j,k} = \sum_{p=1}^{N_i} \frac{Ze}{\Delta V_{i,j,k}} \prod_{d=1}^3 W^{(n)}\left( \frac{ (\textbf{r}_{i,j,k} - \textbf{r}_p) \cdot \hat{\textbf{e}}_d }{\Delta x_d}\right)
    \label{eq:scatter_general}
\end{equation}
and
\begin{equation}
    \textbf{E}(\textbf{r}_p) = \sum_{i,j,k} \textbf{E}_{i,j,k} \prod_{d=1}^3 W^{(n)}\left( \frac{ (\textbf{r}_{i,j,k} - \textbf{r}_p) \cdot \hat{\textbf{e}}_d }{\Delta x_d}\right)
    \label{eq:gather_general}
\end{equation} where $(i,j,k)$ are the indices of an arbitrary node in the domain, $\Delta V_{i,j,k}=\prod_{d=1}^3 \Delta x_d$ is the volume of the cell within the nodes $(i,j,k)$ and $(i+1,j+1,k+1)$, $W^{(n)}$ is the shape function of order $n$, $\textbf{r}_{i,j,k}$ is the position of the node $(i,j,k)$, $\textbf{r}_p$ is the position of the ion $p$ and $\textbf{E}_{i,j,k}$ is the electric field at the node $(i,j,k)$. The implemented shape functions ranged from order 2 to order 6. For instance, the shape function of order 2 is given as follows:
\begin{equation}
  W^{(2)}(x) = 
    \begin{cases}
      1 - |x| & \text{for } 0 \leq x \leq 1, \\
      0 & \text{otherwise.}
    \end{cases}
    \label{eq:W_2}
\end{equation} Through the rest of the paper, a shape function of order 2 is utilized unless otherwise indicated. Shape functions of order 4 and 6 are described in appendix \ref{sec:appendix} \cite{ABE_1986_shape_functions}.

Integration of the equations of motion was conducted using the Verlet algorithm. The Poisson equation solver through a 3D$-$FFT, interpolation operations, and integration of equation of motion were implemented in CUDA$-$C kernels through the open source library Cupy and called from python.

In the first part of this work, shown in section \ref{sec:DIH}, ions are simulated as macroparticles. The number of macroparticles was 5000, and the simulated ion density was $2.5\times 10^{24}$~m$^{-3}$ with a timestep of $\Delta t = 10^{-3} \omega_{pi}^{-1}$. The positions of the macroparticles were initialized with a uniform random distribution and the velocities with a Maxwellian distribution at room temperature. No electron or neutral species were included since the objective of this part is to study strongly coupled ion-ion interactions in atmospheric pressure plasmas~\cite{Acciarri2022}. To justify not including neutral atoms, the ion-neutral collision frequency at large ionization fractions is much smaller than the ion plasma frequency which sets the timescale of DIH \cite{Acciarri2022,Acciarri2023_Diffusion}. In addition, the elevated electron temperature makes electron screening of ions negligible, as observed by Shaffer et. al. \cite{Shaffer_POP_2017}. Thus, electrons are considered as a background non-interacting neutralizing species. Thus, they were not included in the simulation \cite{Acciarri2022,Acciarri2023_Diffusion}. Consequently, ion-ion interactions can be treated as an one-component plasma (OCP).

For the second part of this work, detailed in section \ref{sec:ACH}, PIC was used to simulate the electron dynamics on a timescale given by the electron plasma period in order to study ACH. Hence, ions acted as a background stationary species and were not included in the simulations since only the first electron plasma periods were simulated. For the electron simulations, equation~(\ref{eq:rho}) was modified taking the ion density as uniform and the electron density as variable with the position $n_e(\textbf{r})$, calculated from the interpolation of the macroparticle positions.

\section{Disorder-Induced Heating}\label{sec:DIH}

\subsection{Molecular Dynamics Results}\label{subsec:DIH_MD}

\begin{figure}
    \includegraphics[width=8cm]{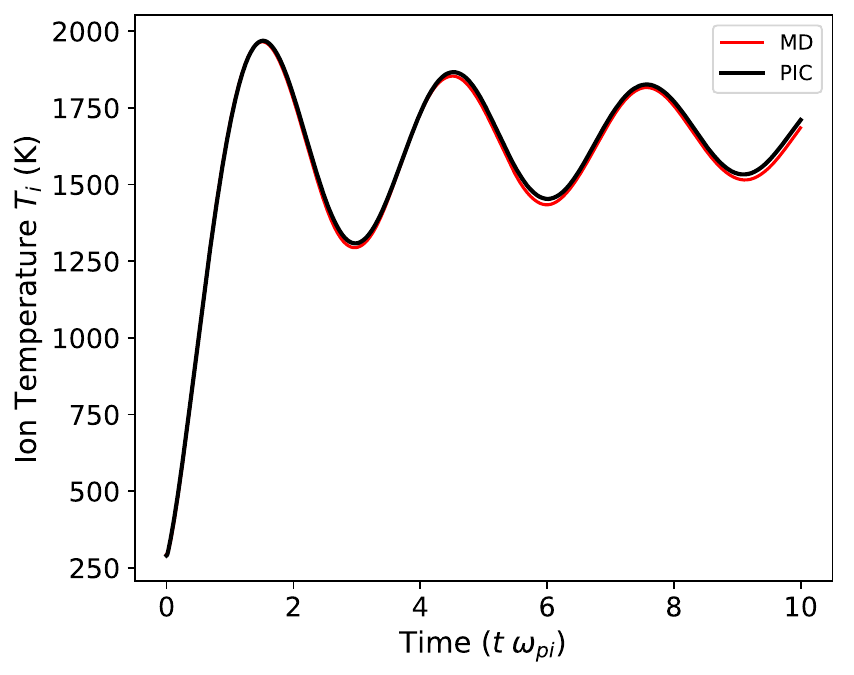}
    \caption{Evolution of the ion temperature obtained with a 3D3V PIC simulation with a grid spacing  of $\Delta x / a_{ii} \approx 0.042$ and from an MD simulation for the same initial conditions and density.}
    \label{fig:Ti_comparison_MD}
\end{figure}

Figure \ref{fig:Ti_comparison_MD} presents the evolution of the ion temperature over the first $10$ ion plasma periods, obtained from the MD simulation described in section \ref{subsec:simulation_setup_MD}. The ion temperature rapidly increases from room temperature to a peak of $\approx 2000$ K during the initial time period of $1.5 \omega_{pi}^{-1}$. 
Throughout the evolution, the ion temperature exhibits pronounced fluctuations, a result of oscillations in the exchange of potential and kinetic energy as the system relaxes to equilibrium at a Coulomb coupling parameter larger than unity~\cite{Acciarri2022,Killian_UCNP}. 
The absence of ion-neutral interactions in the simulation prevents the relaxation of ion and neutral temperatures that was studied in earlier MD simulations~\cite{Acciarri2022}. The ion temperature shown in figure \ref{fig:Ti_comparison_MD} provides a target to obtain with PIC simulations.

After ionization the ions have significant excess potential energy since the interactions between particles changes from a short range (Lennard-Jones) to a long range (Coulomb) potential. Hence, ions move to their lowest potential configuration converting the excess potential energy into kinetic energy. This, combined with an initial Coulomb coupling parameter $\Gamma \approx 12$ results in the temperature increase observed in figure \ref{fig:Ti_comparison_MD}, known as DIH~\cite{Acciarri2022,Killian_UCNP}. This effect has been previously observed in MD simulations of partially ionized atmospheric pressure plasmas and experiments of ultracold neutral plasmas~\cite{Killian_UCNP,Acciarri2022}. After DIH, ions overshoot their equilibrium positions leading to oscillations of the Coulomb potential energy near the ion plasma frequency. Since the total energy is conserved during the MD simulation and ions are strongly coupled after DIH, those oscillations translate to observable fluctuations in ion temperature as shown in figure \ref{fig:Ti_comparison_MD}.

While in this work no ion-neutral interactions are included, in partially ionized plasmas DIH is followed by an ion-neutral temperature relaxation that decreases the ion temperature and increases the neutral gas temperature until equilibrium is reached. The impact of DIH is substantial, with the resulting equilibrium temperature increase reaching several multiples of room temperature in atmospheric pressure plasmas with ionization fractions $x_i \geq 10^{-2}$ (ion densities $n_i \geq 2.5 \times 10^{23}$ m$^{-3}$)~\cite{Acciarri2022}. To illustrate the effect of DIH, a model was presented in Ref.~\onlinecite{Acciarri2022}, delineating the calculation of equilibrium temperature based on energy conservation principles. The model for the equilibrium temperature is
\begin{equation}
T_\textrm{eq} = x_i  T^{\max}_i + (1-x_i)  T_n(t=0)
\label{eq:T_eq}
\end{equation} where $T_n(t=0)$ is the neutral gas temperature before ionization, $T^{\max}_i$ is the maximum ion temperature due to DIH and $x_i$ is the ionization fraction. 
Equation~(\ref{eq:T_eq}) has been validated through its agreement with MD simulations of discharges at atmospheric pressure~\cite{Acciarri2022}. The maximum ion temperature due to DIH accounts for the change in the Coulomb potential energy
\begin{equation}
T^{\max}_i = \frac{1}{1.91} \frac{Z^2e^2}{4\pi\epsilon_0k_B} \frac{1}{a_{ii}},
\label{eq:Ti_max}
\end{equation} where 1.91 is a numerical factor obtained from MD simulations \cite{Acciarri2022}. Future work will clarify the need of this numerical factor and its dependence on the particle density.

\subsection{PIC: Influence of Grid Resolution}\label{subsec:DIH_PIC_grid_resolution}

\begin{figure}
    \includegraphics[width=8cm]{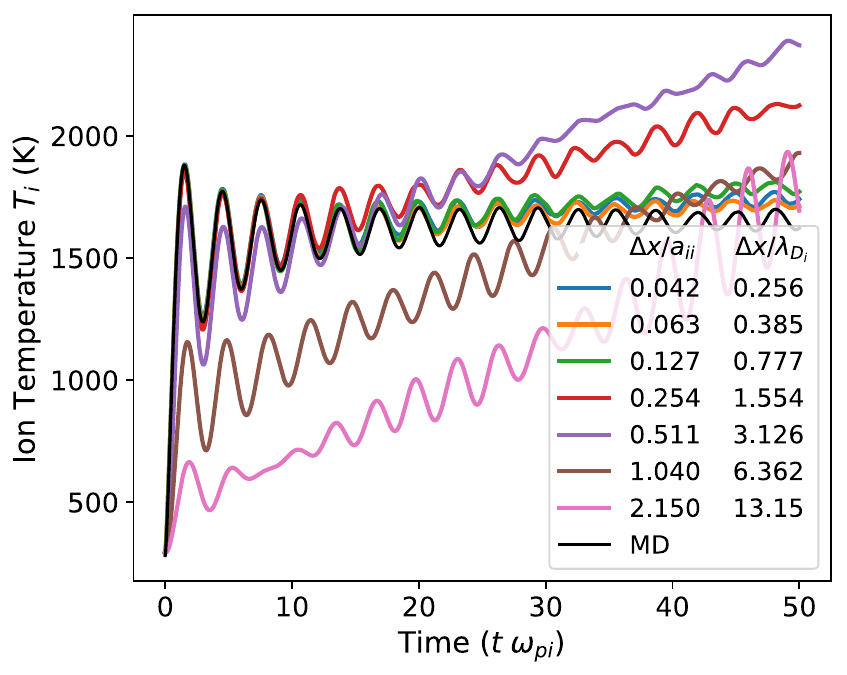}
    \caption{Evolution of the ion temperature for an ion density of $n_i=2.5\times 10^{24}$~m$^{-3}$ and initial room temperature obtained from MD and using the PIC method for different grid spacing.  }
    \label{fig:Ti_convergence_test_dx_aii_spwt_1}
\end{figure}

Figure \ref{fig:Ti_convergence_test_dx_aii_spwt_1} presents the evolution of ion temperature from initial room temperature and a random distribution of positions for different uniform grid spacing $\Delta x$ and unity macroparticle weight. At the simulated density, the expected maximum ion temperature after $t \omega_{pi} = 1.5$ is approximately $1800$ K from equation~(\ref{eq:Ti_max}). The results demonstrate that when $\Delta x / a_{ii}>1$ and $\Delta x / \lambda_{D_i}>1$, where $\lambda_{D_i}$ is the ion Debye length, DIH is not fully observed, nor is PIC heating absent. 
Here, DIH corresponds to the rapid initial temperature increase over $\sim1.5 \omega_{pi}^{-1}$, while PIC heating is the longer-time linear increase over tens of $\omega_{pi}^{-1}$.
If $\Delta x / a_{ii}<1$ and $\Delta x / \lambda_{D_i}>1$, the temperature increase due to DIH aligns more closely with the expected value, but PIC heating still persists. Ultimately, when both $\lambda_{D_i}$ and $a_{ii}$ are resolved, DIH is fully observed and there is an agreement with the expected temperature increase. Furthermore, the evolution of the ion temperature for the smallest grid spacing of $\Delta x \approx 0.042 a_{ii}$ shows a good agreement with the target evolution obtained with MD over the $~10 \omega_{pi}^{-1}$ timescale, as shown in figure \ref{fig:Ti_comparison_MD}.

\begin{figure}
    \includegraphics[width=8cm]{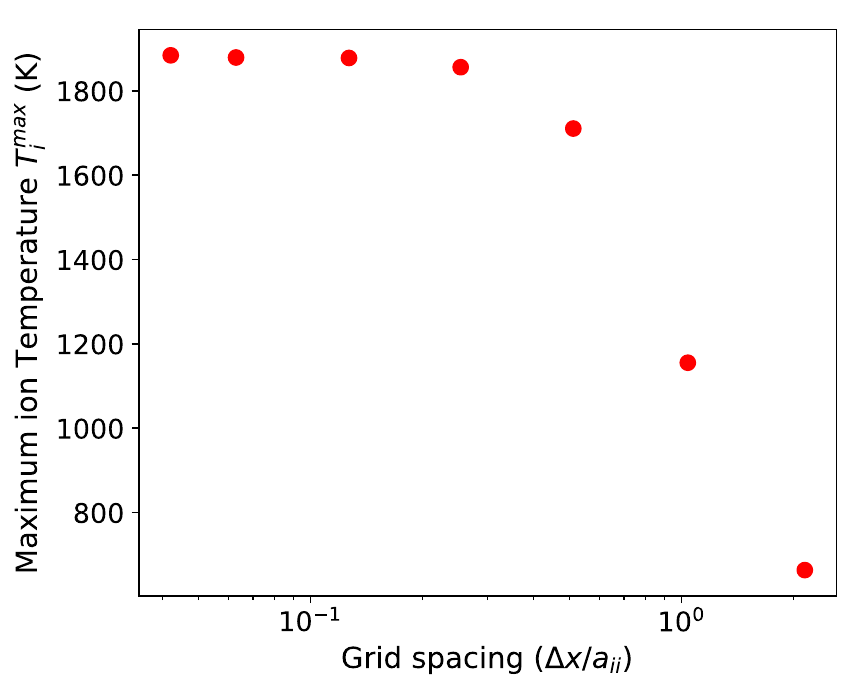}
    \caption{Ion temperature of the first peak (at $t \approx 1.5\;\omega_{pi}^{-1}$) for different grid spacing values  $\Delta x /  a_{ii}$. The maximum ion temperature converges to the expected physical value for $\Delta x / a_{ii} \leq 0.1$.}
    \label{fig:Ti_max_convergence_test_dx_aii_spwt_1}
\end{figure}

Figure \ref{fig:Ti_max_convergence_test_dx_aii_spwt_1} shows the maximum ion temperature, taken from the first peak of each simulation, for different grid spacing. It is clear that in order to fully capture DIH with a PIC simulation, the grid spacing must be $\Delta x \lesssim 0.1 a_{ii}$. It is important to underscore that in strongly coupled plasmas, the average interparticle spacing is larger than the Debye length. Thus, resolving the Debye length to avoid PIC heating requires resolving the average interparticle spacing. This implies that, on average, there are fewer particles in the simulation than grid cells, contradicting the standard practice in PIC simulations where multiple macroparticles per cell are essential to mitigate statistical noise. As an example, for a grid resolution of $\Delta x = 0.1 a_{ii}$ and the simulated density, there is an average of $2.38 \times 10^{-4}$ macroparticles per cell. This contrasts with the normal operation of PIC simulations of weakly coupled plasmas, where $\lambda_{D_{i}} \gg a_{ii}$ and the Debye length can be resolved while having multiple macroparticles per cell.

\subsection{PIC: Influence of Macroparticle Weight}\label{subsec:DIH_PIC_weight}

\begin{figure}
    \includegraphics[width=8cm]{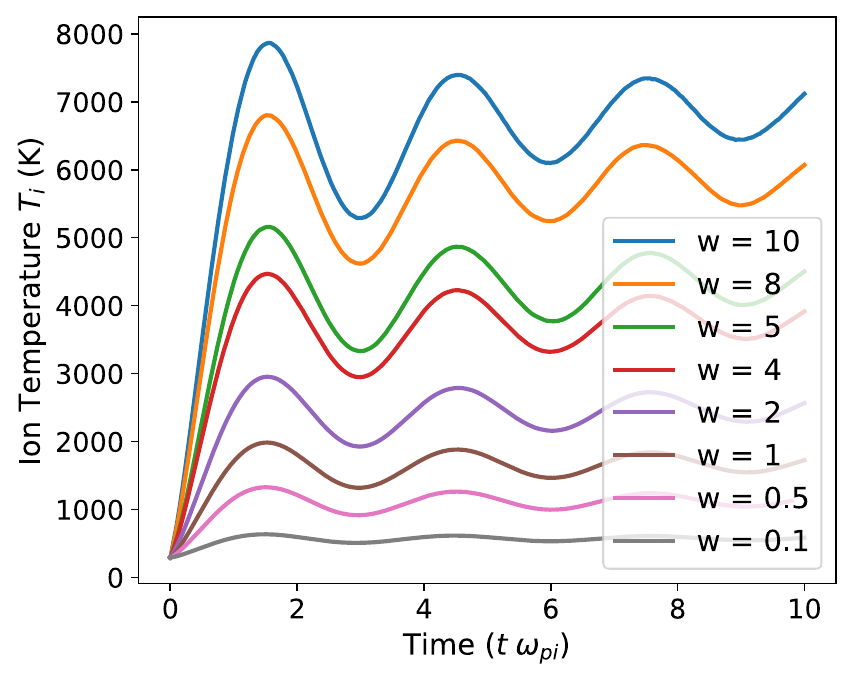}
    \caption{Evolution of the ion temperature using a grid spacing of $\Delta x / a_{ii}\approx 0.042$ for different macroparticle weights $w$.}
    \label{fig:Ti_t_linear_spwt_comparison}
\end{figure}

\begin{figure}
    \includegraphics[width=8cm]{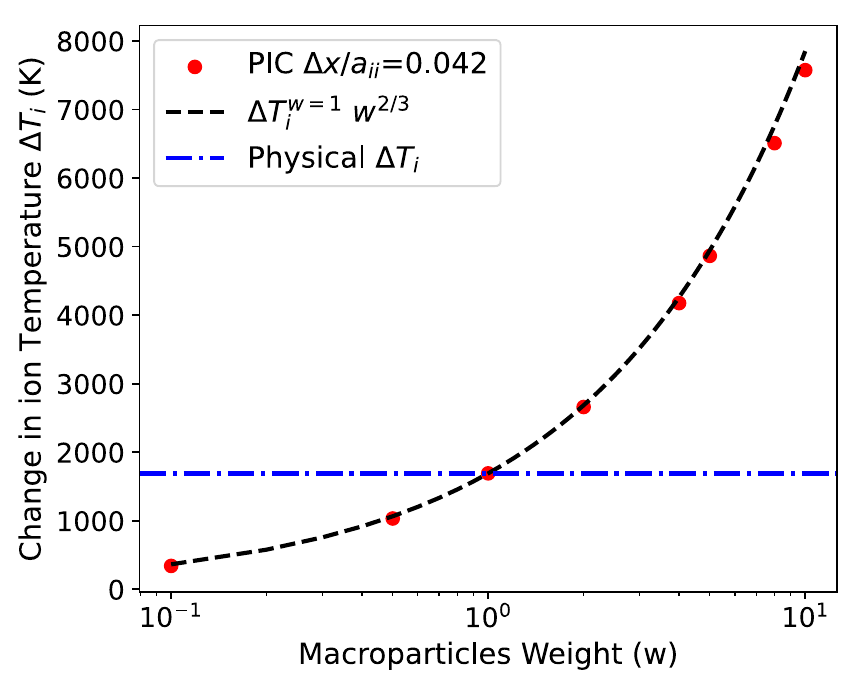}
    \caption{Change in the ion temperature obtained with PIC simulations at different macroparticle weights. The change in temperature obtained with the scaling law $\Delta T_i^w = w^{2/3} \Delta T_i$ is shown for comparison.}
    \label{fig:dTi_vs_spwt_log}
\end{figure}

Figure \ref{fig:Ti_t_linear_spwt_comparison} presents the evolution of ion temperature for different macroparticle weights $w$, using a grid spacing of $\Delta x / a_{ii}\approx 0.042$. 
When $w > 1$ the observed ion temperature exceeds the expected physical value. Conversely, when $w < 1$, the observed temperature is lower. This shift in ion temperature with varying $w$ becomes more pronounced as $w$ is further increased or decreased from unity. The effect of the macroparticle weight on the change in the ion temperature, from room to equilibrium, is shown in figure \ref{fig:dTi_vs_spwt_log}. This highlights the tendency that temperature artificially increases rapidly with macroparticle weight. 

\begin{figure}
    \includegraphics[width=8.5cm]{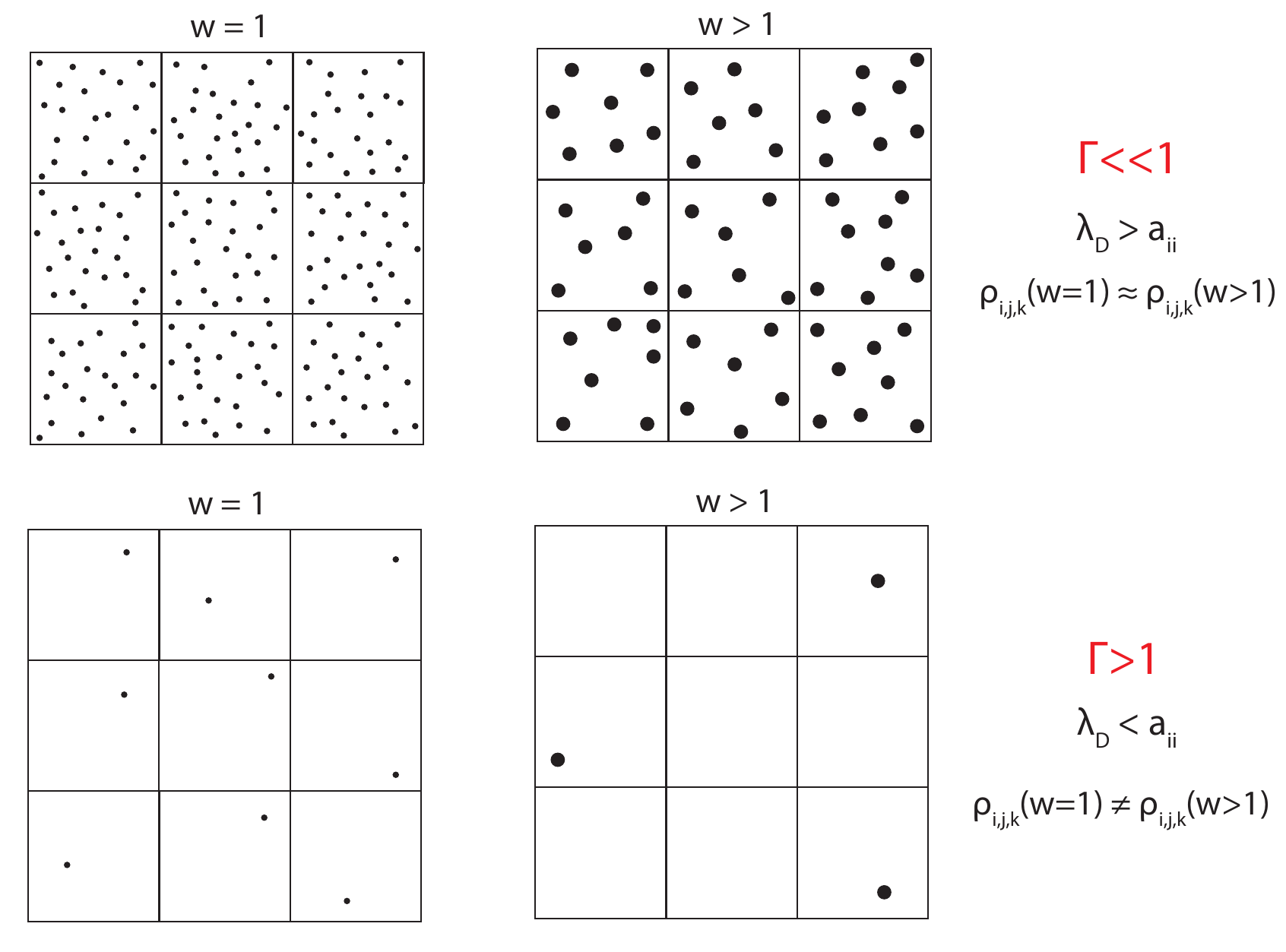}
    \caption{Illustration of the distribution of macroparticles in PIC simulations for (top) weakly vs. (bottom) strongly coupled plasmas assuming a grid spacing of approximately the Debye length.}
    \label{fig:weakly_vs_strongly_coupled_diagram}
\end{figure}

Figure \ref{fig:weakly_vs_strongly_coupled_diagram} aims to provide a  qualitative illustration of this influence. In standard PIC simulations of weakly coupled plasmas, the ion Debye length is much larger than the average interparticle spacing. Therefore, a large number of macroparticles is included within each cell and the resulting charge density aligns closely with the actual physical density in the computational grid. However, this is not the case in strongly coupled plasmas. Here, maintaining an average of fewer than one macroparticle per cell ($\Delta x \leq 0.1 a_{ii}$) is vital for accurately resolving the strong ion-ion correlations and for obtaining a precise representation of DIH. For the same increase in particle weight, a strongly coupled simulation suffers a greater change in the density across adjacent cells when compared to a weakly coupled simulation.  Raising the macroparticle weight above one results in a numerical localization of the charge density, which locally augments the electric field and subsequently elevates the overall potential energy. Conversely, lowering the macroparticle weight below one incurs the opposite effect, smoothing the charge density artificially resulting in the lower DIH observed in figure \ref{fig:Ti_t_linear_spwt_comparison}.

Next, we consider a quantitative description of the influence of the macroparticle weight on the ion temperature. The potential energy of ions in an OCP at equilibrium can be expressed as \cite{HANSEN201313}
\begin{equation}
    \textrm{PE} = 2 \pi n_i N_i \int^\infty_0 dr \; r^2 \phi(r) g(r), 
    \label{eq:PE_definition}
\end{equation} where $n_i$ is the ion density, $N_i$ the number of ions and $\phi(r)$ is the interaction potential. Here, $g(r)$ is the ion pair correlation function, which is defined by setting $n_i g(r) 4 \pi r^2 dr = N_i(r)$, where $N_i(r)$ is the total number of ions in a spherical shell of radius $r$ and thickness $dr$ centered on a chosen particle. 
It is a density profile of other particles referenced to a particle at the origin, normalized to the background density. 
If there is a change in the potential energy between two configurations, the corresponding change in the ion temperature is
\begin{equation}
    \Delta T_i = -  \frac{\Delta \textrm{PE}}{\frac{3}{2}k_B N_i},
    \label{eq:Delta_Ti_definition}
\end{equation} where $T_i$ is the ion temperature and $k_B$ is the Boltzmann constant. Combining expressions (\ref{eq:PE_definition})-(\ref{eq:Delta_Ti_definition}) and scaling the radial distance using the average interparticle distance between ions, the change in the ion temperature from a randomly distributed configuration in space to a final equilibrium state characterized by the pair correlation function $g(r)$ is expressed as
\begin{equation}
    \Delta T_i = \frac{1}{3} \frac{n_i q_i^2 a_{ii}^2}{k_B \epsilon_0} \int^\infty_0 d \tilde{r} \; \tilde{r} [1- g(\tilde{r})],
    \label{eq:Delta_Ti}
\end{equation} where the pair correlation function of the initial randomly distributed configuration is taken as $1$ at any radial distance and $\tilde{r} = r/a_{ii}$. Equation~(\ref{eq:Delta_Ti}) can be used to estimate the change in temperature when the macroparticle weight is $w=1$. For the more general case, $w \neq 1$, the corresponding scaling laws $n_i^w \rightarrow n_i/w$, $q_i^w\rightarrow q_i\;w$ and $a_{ii}^w\rightarrow a_{ii}\;w^{1/3}$ must be replaced in equation~(\ref{eq:Delta_Ti}). The expected change in temperature for an arbitrary macroparticle weight can be expressed as
\begin{equation}
    \Delta T_i^w = \frac{1}{3} \frac{n_i q_i^2 a_{ii}^2}{k_B \epsilon_0} w^{5/3} \int^\infty_0 d \tilde{r} \; \tilde{r} [1-g(\tilde{r})],
    \label{eq:Delta_Ti_w}
\end{equation} where $n_i$, $q_i$ and $a_{ii}$ are the physical values of the density, charge and average interparticle spacing respectively. 
Since the macroparticle temperature is the physical temperature times the macroparticle weight [$T_i^w = T_i w$], the change of the physical ion temperature is predicted to scale proportionally to $w^{2/3}$: $\Delta T_i \propto w^{2/3}$. This means that for macroparticle weights $w \neq 1$, the change in temperature due to DIH computed from PIC will differ from the physical value by a factor of $w^{2/3}$, assuming that the interparticle spacing of macroparticles is well resolved.

\begin{figure}
    \includegraphics[width=8cm]{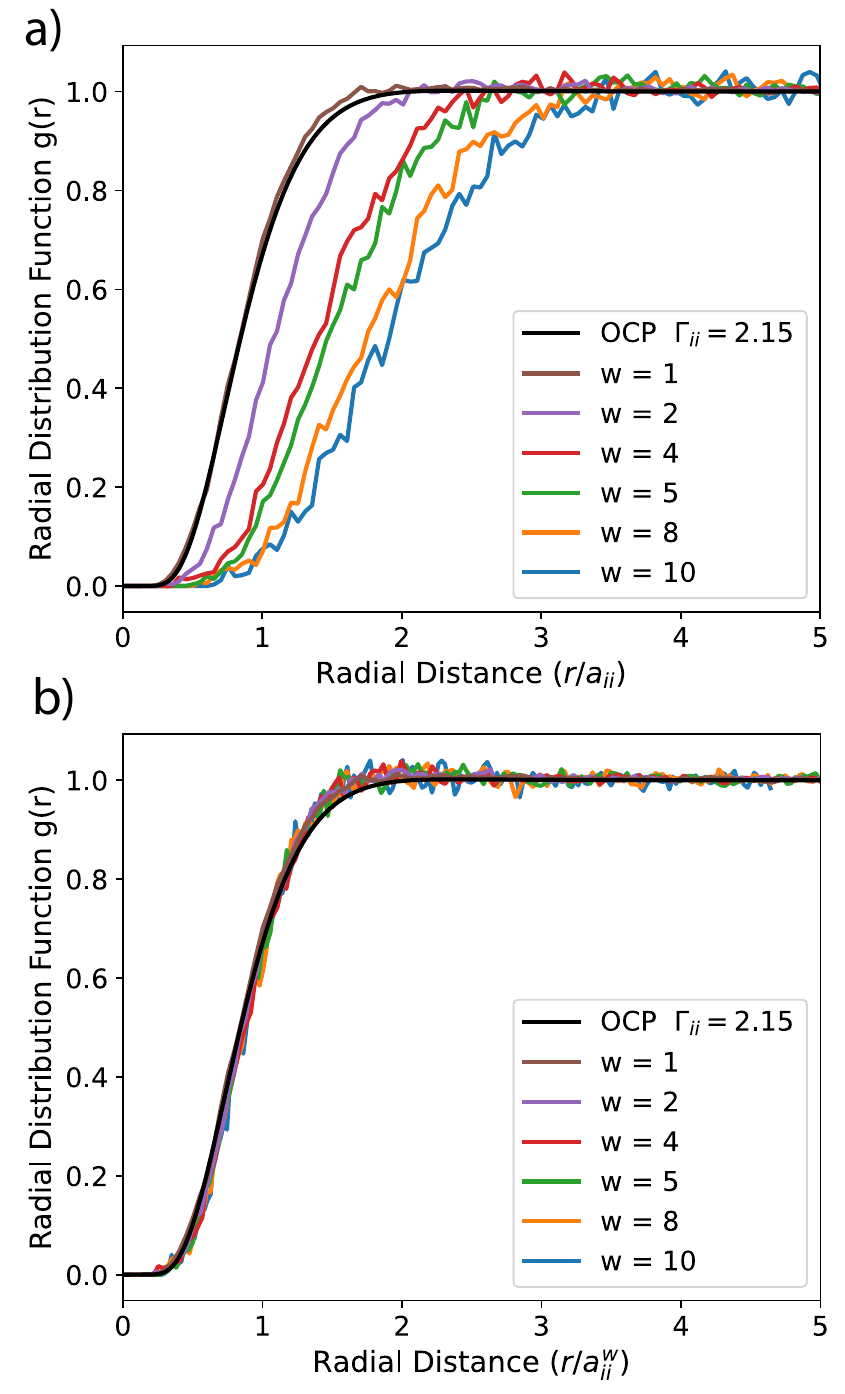}
    \caption{Pair correlation function $g(r)$ for different macroparticle weights at equilibrium taking as reference the physical average interparticle spacing (a) and taking as reference the average macroparticle spacing $a_{ii}^w$ (b). The pair correlation function of an OCP at the equilibrium $\Gamma_{ii}$ is shown for comparison.}
    \label{fig:gr_spwt_together}
\end{figure}

Figure \ref{fig:gr_spwt_together}(a) displays the pair correlation function $g(r)$, obtained by averaging over all macroparticles between the time intervals $t \omega_{pi} = 9.5$ and $t \omega_{pi} = 10$, for different weights. The pair correlation function of an OCP at the same equilibrium temperature is included for comparison. As expected, the pair correlation function corresponding to a macroparticle weight $w=1$ aligns well with the OCP $g(r)$ since individual physical particles are simulated. However, for $w \neq 1$, the pair correlation functions appear shifted relative to the OCP function. Not surprisingly, when the radial distance is scaled by the corresponding average macroparticle distance $a_{ii}^w$, all $g(r)$ values align with the OCP $g(r)$ at the equilibrium temperature, as shown in figure \ref{fig:gr_spwt_together}(b).

The scaling law $\Delta T_i^w = w^{2/3} \Delta T_i$ demonstrates excellent alignment with the increase in the ion temperature obtained from PIC simulations across various macroparticle weights, as exhibited in figure~\ref{fig:dTi_vs_spwt_log}. While employing macroparticles is standard in traditional PIC simulations, these findings suggest that in strongly coupled plasmas the weight of macroparticles significantly impacts potential energy and consequently the kinetic energy and overall plasma dynamics. This effect does not physically manifest in weakly coupled plasmas, where the average Coulomb potential energy is significantly less than the kinetic energy ($\Gamma_{ii}\ll 1$), thereby facilitating the use of macroparticles.

The main takeaway of these results is that in order to capture DIH and, generally, strong correlations with PIC simulations it is necessary to resolve the average interparticle spacing, combined with a unity macroparticle weight. Furthermore, the need to resolve such a small spatial scale adds an extra constraint to the timestep, since macroparticles should not travel a distance larger than a cell per timestep. These additional constraints make traditional PIC simulations intractable for atmospheric pressure plasmas due to the prohibitive associated computational cost. To illustrate, consider a simulation run at atmospheric pressure with an ionization fraction of $10\%$, and $10^6$ macroparticles of unity weight. In this scenario, the simulation volume would only span an approximate length of $736$ nm. Contrastingly, a simulation run under similar conditions of ionization fraction and total macroparticle count, but at a lower pressure of 1 mTorr and with a macroparticle weight of $10^7$, would encompass a simulation volume of approximately $14$ cm - akin to the length scale of a plasma device. These discrepancies in scale underscore the limitations of PIC simulations in capturing the behavior of strongly coupled plasmas at atmospheric pressure, urging us to explore alternative computational approaches in future work. Moreover, if a fluid approach is used to model ions at atmospheric pressure and conditions relevant to the strongly coupled regime, effects such as DIH or changes in diffusion rates will be completely missed and the neutral gas temperature could be mistakenly underestimated, making the results from a simulation highly questionable due to important physics being ignored. A possible solution would be to use the P3M method instead of traditional PIC, however to capture the correct change in the Coulomb potential the particle-particle part of P3M would require ions to have a unity macroparticle weight. While this would reduce the computational cost associated to the grid resolution, it would still make simulations of device-scale plasmas intractable due to the unity macroparticle weight.

\subsection{PIC Heating and Influence of Interpolation Scheme }\label{subsec:DIH_PIC_heating_interpolation}

In PIC simulations, it is crucial that the grid resolution is sufficiently refined to accurately resolve the Debye length. Failure to meet this requirement can yield numerical artifacts and inaccurate simulation outcomes \cite{Birdsall}. A well-known artifact is the numerical phenomenon of ``PIC heating'', leading to an overestimation of the plasma temperature \cite{Chacon_2020}. This artifact emerges due to the aliasing of interpolation errors that occur between the mesh quantities and macroparticles. Figure \ref{fig:Ti_convergence_test_dx_aii_spwt_1} displays the manifestation of PIC heating in simulations where the ion Debye length is inadequately resolved. It is noteworthy that in strongly coupled plasmas, where $\lambda_{D_i}< a_{ii}$, avoiding PIC heating by resolving the Debye length implicitly ensures the resolution of the average interparticle spacing $-$ a prerequisite identified in this study.

\begin{figure}
    \centering
    \includegraphics[width=8
cm]{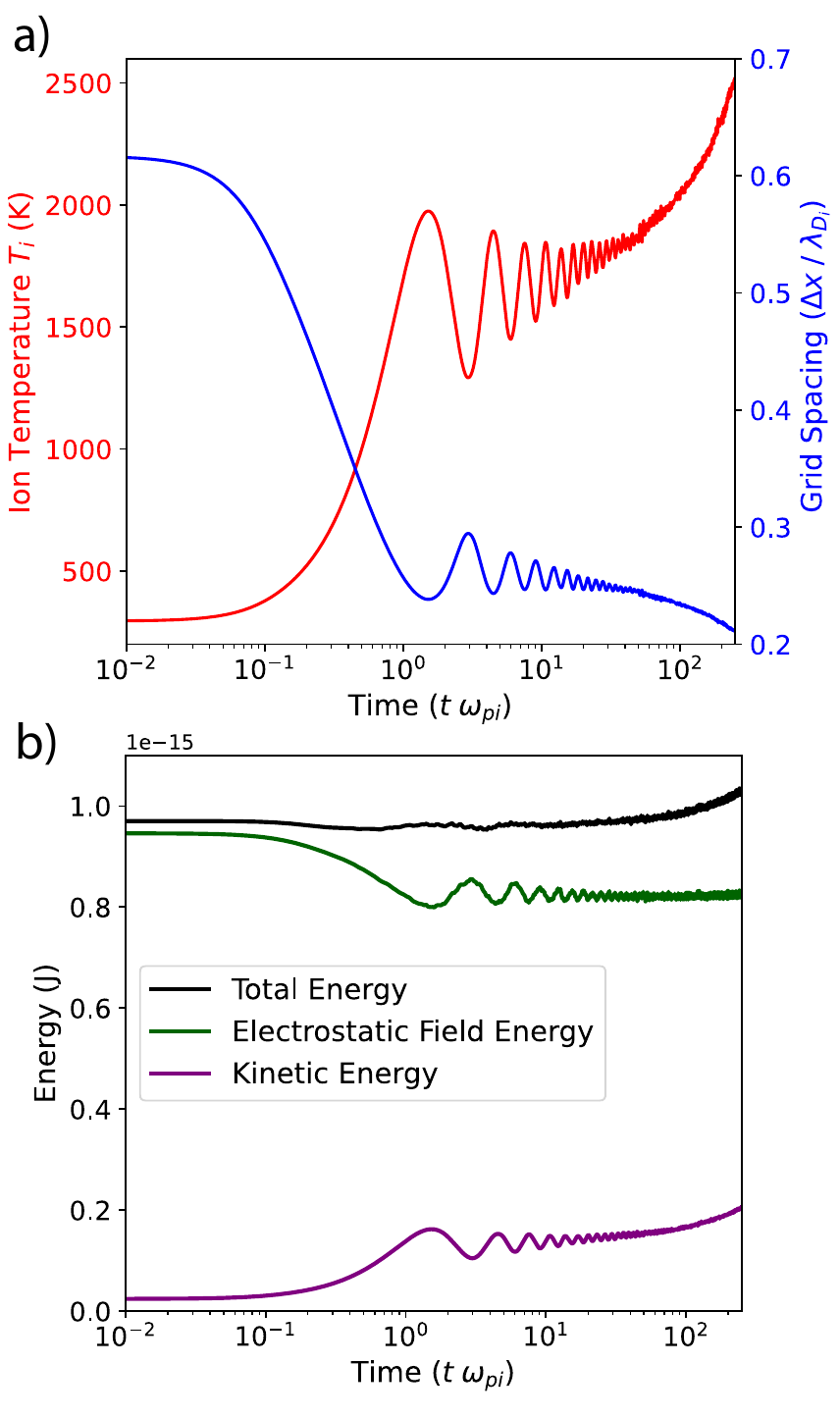}
    \caption{(a) Evolution of the ion temperature and grid spacing $\Delta x / \lambda_{D_i}$ for an ion density of $2.5\times10^{24}$ m$^{-3}$, unity macroparticle weight and average number of macroparticles per cell $N_c=0.00023$. (b) Evolution of the total, electrostatic and kinetic energies over time.}
    \label{fig:long_PIC_heating}
\end{figure}

Resolving the smallest Debye length eradicates PIC heating in dilute plasmas. However, in the strongly coupled regime, PIC heating persists. When a PIC simulation of a strongly coupled plasma resolves both the Debye length and the average interparticle distance, there is, on average, less than one macroparticle per cell. This leads to a discontinuity in the electric field every time a particle crosses a cell boundary, thereby inducing PIC heating. Therefore, even when all relevant physical distances are correctly resolved, PIC heating still occurs on a much longer timescale—hundreds to thousands of plasma periods. This effect is demonstrated in figure \ref{fig:long_PIC_heating} (a), where at the onset of a simulation, $\Delta x \approx 0.6 \; \lambda_{D_i}$. However, after DIH, the plasma continues to experience an increase in temperature due to PIC heating on a timescale of hundreds to thousands of plasma periods. The difference in timescales between DIH and PIC heating is illustrated in figure \ref{fig:long_PIC_heating} (b). The total energy is conserved for the first several plasma periods, where the increase in kinetic energy occurs due to a decrease in the electrostatic field energy. On a longer timescale, spanning 100s of plasma periods, the lowest potential configuration is maintained but the kinetic energy, and therefore total energy, increase due to PIC heating.

Recent research has explored strategies to decrease the computational cost and focus on macroscopic physical phenomena. These strategies involve reducing the PIC heating growth rate, while under resolving the Debye length through advanced energy conserving integration schemes, high-order macroparticle shape functions, and filtering methods \cite{Chacon_2020,Chacon_2011,MARKIDIS_2011,Chacon_2013,Chacon_2014,Chacon_2015,KNOLL_2004_Nwton_Krylov_method}. These strategies, in particular high order shape functions and filtering methods, rely on the plasma being weakly coupled. While these approaches may reduce the growth rate of PIC heating, it is found that it further reduces DIH. 

\begin{figure}
    \centering
    \includegraphics[width=7.9
cm]{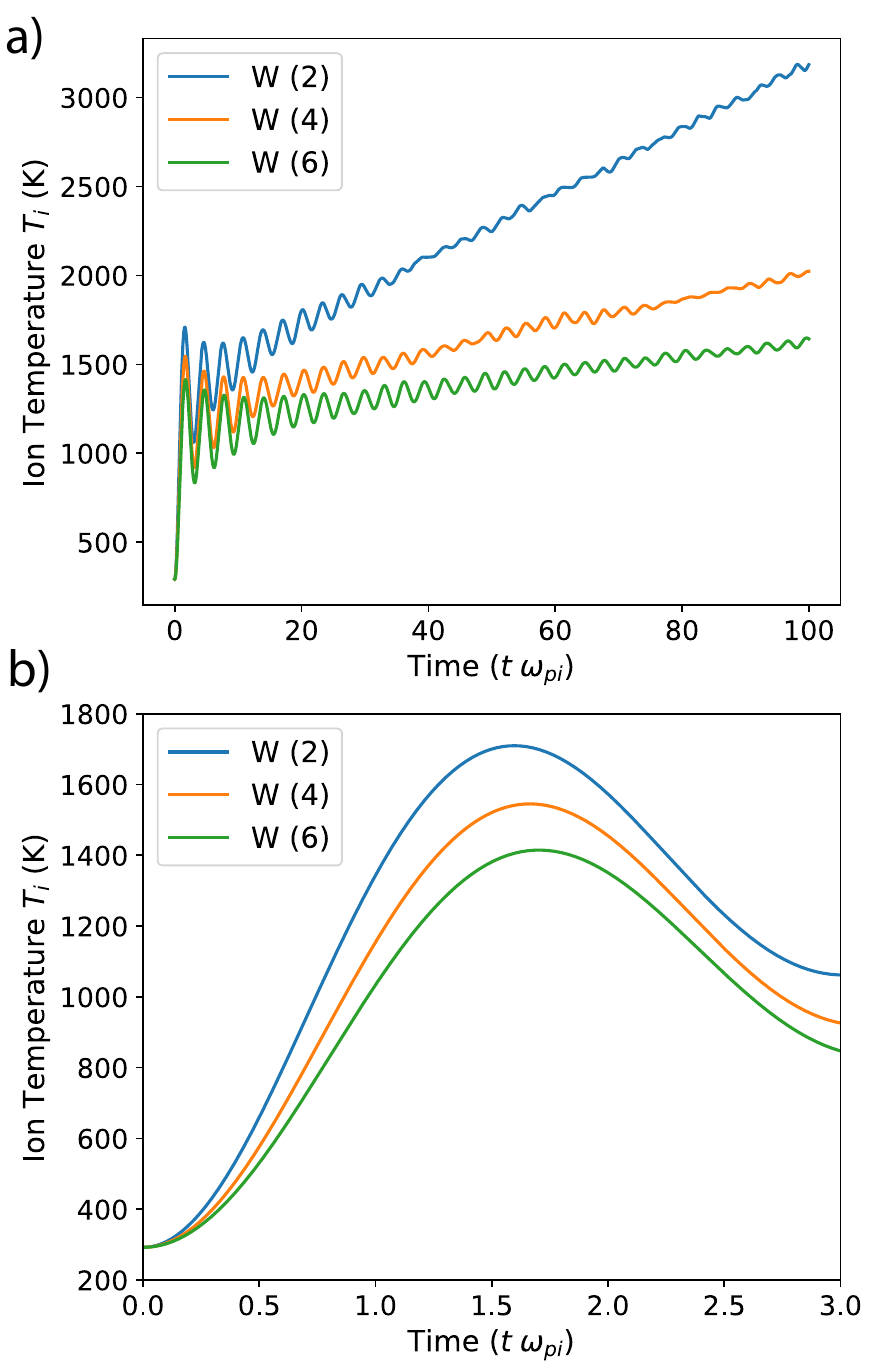}
    \caption{(a) Evolution of the ion temperature obtained from PIC simulations with different shape functions for the first 100 ion plasma periods, and (b) the first 3 plasma periods.  The grid resolution is $\Delta x / a_{ii} \approx 0.51$ and $\Delta x / \lambda_{Di} \approx 3.13$ and the ion density is $2.5\times10^{24}$ m$^{-3}$. Increasing the order of the shape function reduces the growth rate of the observed PIC heating and DIH.}
    \label{fig:Verlet_PIC_heating_together}
\end{figure}

Figure \ref{fig:Verlet_PIC_heating_together} a) shows the evolution of the ion temperature in a simulation where the grid resolution ($\Delta x / a_{ii} \approx 0.51$ and $\Delta x / \lambda_{D_i} \approx 3.13$) under resolves the ion Debye length, across different shape functions ranging from order 2 to order 6, as described in equations (\ref{eq:W_2}), (\ref{eq:W_4}) and (\ref{eq:W_6}). Consistent with existing literature, the results demonstrate that the growth rate of PIC heating significantly decreases with higher order shape functions. However, in a system exhibiting strong ion-ion correlations, it is crucial to resolve the average interparticle spacing, as delineated in this study. This necessity results in an average of less than one particle per cell. Within such a framework, employing high order shape functions can result in the delocalization of the charge density, leading to an artificial reduction in the effective Coulomb coupling parameter of macroparticles. This, in turn, produces an underestimation of the electric field, which subsequently reduces the observed disorder-induced heating, as illustrated in figure \ref{fig:Verlet_PIC_heating_together} (b). This finding underscores that due to the inherent nature of strong Coulomb coupling, the charge density must remain numerically unchanged, both locally and globally to accurately represent the physical state of the plasma. This further limits the applicability of PIC simulations for plasmas where ion-ion interactions are strongly correlated.

\subsection{Reduced Dimensions}\label{subsec:DIH_reduced_dimensions}

Running PIC simulations in reduced spatial dimensions is a common approach for systems with some kind of spatial symmetry. While this approach significantly reduces the computational cost, physical quantities such as charge and particle densities, are based on volumetric properties. The usual approximation to connect a physical volumetric density with a numerical density in reduced dimensions is to consider a constant area (or length) perpendicular to the simulation domain, allowing the calculation of a volume for each cell \cite{Donko_2021_eduPIC}. This approach, however, is equivalent to projecting the positions of all particles within the physical volume onto an element in the corresponding reduced dimension. While this method is acceptable in the weakly coupled regime, where interactions between individual particles are ignored and multiple macroparticles are located within each cell, it artificially reduces the interparticle distance. If a plasma is strongly coupled, that interparticle distance needs to be resolved in order to accurately capture the strong correlations. This, combined with the requirement of unity macroparticle weight described in this work, makes PIC simulations with reduced dimensions not applicable for strongly coupled plasmas.

To illustrate, for an ionization fraction of $10\%$ at atmospheric pressure the ion density is $2.5\times10^{24}$~m$^{-3}$ and the corresponding average interparticle spacing in a 3D physical domain is $a_{ii} = (3/4\pi n_i)^{1/3}=4.57 \times 10^{-9}$~m. However, if a 1D domain is used with unity macroparticle weight, the average interparticle distance in the simulation scales as $a_{ii}^{1D} = 1/n_i = 4 \times 10^{-25}$~m which is 16 orders of magnitude smaller and completely unphysical. Hence, it is clear that scaling down a problem to lower dimensions significantly affects the average interparticle spacing. This exacerbates the issues associated with applying PIC to strongly coupled plasmas.

\section{Artificial Correlation Heating}\label{sec:ACH}

In a PIC simulation, the typical use of macroparticles can alter the coupling strength for electron and ion species. This is observed when the effective coupling strength is artificially enhanced by a macroparticle weight $w>1$, for example as detailed in section \ref{subsec:DIH_PIC_weight} and shown in figure \ref{fig:Ti_t_linear_spwt_comparison}. Consequently, a weakly coupled physical charged species can artificially transition to a strongly coupled macroparticle species. This condition, in conjunction with an average of less than one macroparticle per cell (a circumstance plausible at high plasma densities, where the Debye length is significantly decreased) instigates a novel numerical heating mechanism. This heating mechanism, similar to DIH but distinctively attributed to the macroparticle weight, is henceforth referred to as Artificial Correlation Heating (ACH).

\subsection{Macroparticle Coupling Strength}\label{subsec:ACH_gamma}

When macroparticles are used, the numerical density, charge, temperature and average interparticle distance of macroparticles scale as $n^w_e \rightarrow n_e/w$, $q_e^w\rightarrow e \; w$, $T_e^w\rightarrow T_e\;w$ and $a_{ee}^w\rightarrow a_{ee}\;w^{1/3}$ respectively, where $n_e$, $e$, $T_e$ and $a_{ee}$ are the physical quantities for electron species and $w$ is the macroparticle weight. To clarify, the physical temperature is an intensive quantity and it does not depend on the macroparticle weight, however due to the scaling of the macroparticle kinetic energy with the mass, $T_e^w$ is proportional to $w$ \cite{Birdsall}. Replacing these scaling laws in the equation for the Coulomb coupling parameter for electrons (\ref{eq:gamma}) we have
\begin{equation}
\Gamma_{ee}^w = \frac{e^2}{4 \pi \epsilon_0 a_{ee}} \frac{w^{2/3}}{k_B T_e} = \Gamma_{ee} \; w^{2/3},
\label{eq:gamma_w}
\end{equation} where $\Gamma_{ee}$ represents the physical Coulomb coupling parameter of electrons. The $w^{2/3}$ factor indicates that when the macroparticle weight is significantly larger than unity, it could considerably increase the effective coupling strength between macroparticles. This in turn could artificially enhance the correlations between macroparticle electrons. Consequently, a plasma that is physically weakly coupled could be misrepresented as strongly coupled in a PIC simulation.

\begin{figure}
    \centering
    \includegraphics[width=8cm]{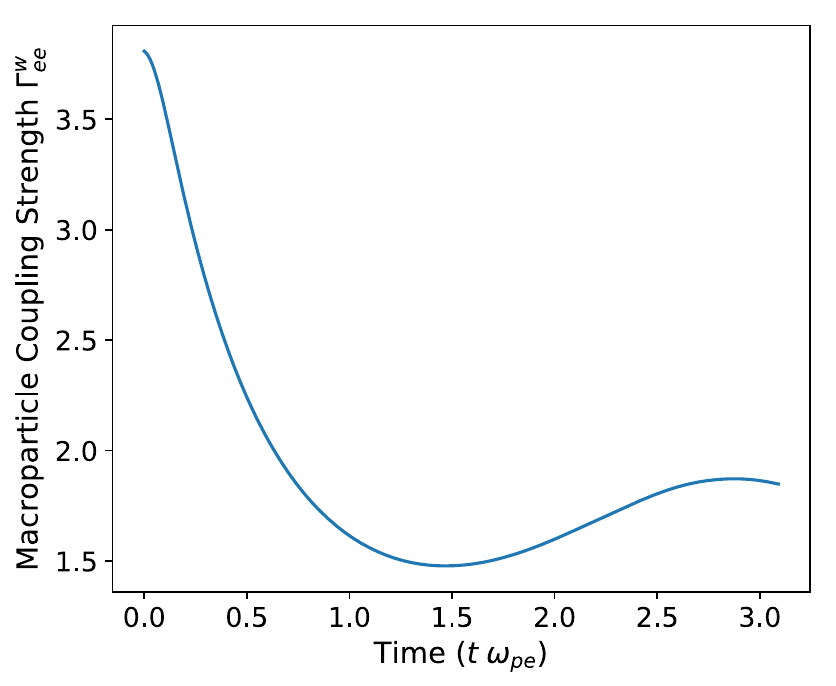}
    \caption{Evolution of the macroparticle coupling strength between electrons in a PIC simulation where ACH is observed. The macroparticle $\Gamma_{ee}^w$ decreases to a value of approximately 1.5 after $1.5 \omega_{pe}^{-1}$. A grid resolution of 0.3$\lambda_{De}$ was used.}
    \label{fig:gamma_w_vs_t_example_ACH}
\end{figure}

To illustrate, consider a plasma with an electron density of $1.25\times 10^{23}$~m$^{-3}$ and an electron temperature of 3~eV. The electron Coulomb coupling parameter in this scenario is approximately 0.0387 which corresponds to the weakly coupled regime. If a PIC simulation is conducted with a timestep of 0.01 of the electron plasma period, a grid resolution of 0.3 the electron Debye length, and a moderate macroparticle weight $w$ of 1000, electrons heat up significantly to a maximum temperature of 7.65~eV within $1.5 \; \omega_{pe}^{-1}$ due to artificial correlation heating, as shown in figure \ref{fig:ACH_Te_vs_t}. This nonphysical rise in the electron temperature is a consequence of a macroparticle coupling strength of approximately 3.87 (derived from equation \ref{eq:gamma_w}), signifying strong coupling and thereby illustrating the potential impact of ACH. This is better observed in figure \ref{fig:gamma_w_vs_t_example_ACH}, where the evolution of the effective coupling strength between macroparticles decreases to a value of approximately 1.5 after 1.5 electron plasma periods.  It is important to underscore that in a simulation with electron impact ionization collisions this process is potentially unstable, since an artificial heating of electrons could result in additional ionization that further increases the electron density and therefore inducing a further rise in the electron temperature due to ACH. This creates the potential for a runaway heating process. 

\subsection{Parameter Space}\label{subsec:ACH_parameter_space}

\begin{figure}
    \includegraphics[width=8cm]{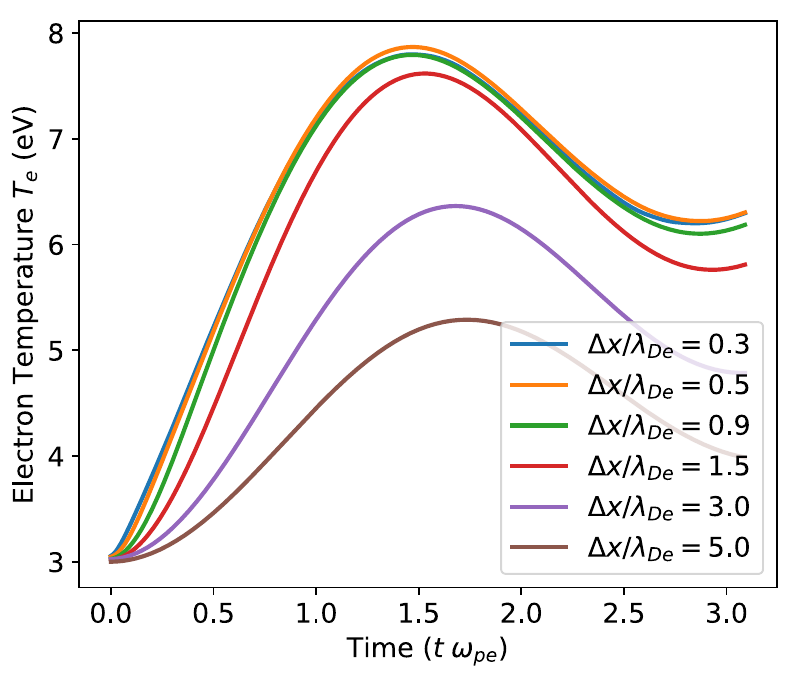}
    \caption{Evolution of the electron temperature for the first few electron plasma periods for different grid spacing. The electron density is $1.25\times 10^{23}$ m$^{-3}$ and the macroparticle weight is 1000.}
    \label{fig:ACH_Te_vs_t}
\end{figure}

Artificial correlation heating is observed not only because of a large, artificially enhanced, macroparticle coupling strength but also, a small number of macroparticles per cell. Hence, the artificially increased correlations between macroparticles are resolved by the grid. This is illustrated in figure \ref{fig:ACH_Te_vs_t}, which shows that as the grid resolution increases the maximum temperature due to ACH increases accordingly with a smaller average number of macroparticles per cell. The number of macroparticles per cell for each of these cases is shown in table \ref{tab:table_ACH}. However, given an electron density and temperature, it is not possible to choose the grid resolution, macroparticle weight and number of macroparticles per cell independently. This is, the number of macroparticles per cell is an immediate consequence of the values chosen for the numerical parameters $(\Delta x / \lambda_{D_e}, w)$ given the physical values of $(n_e,T_e)$. Hence, in scenarios where the combination of small number of macroparticles per cell and large macroparticle weight can happen, combined with a large electron density, ACH could significantly increase the electron temperature. 

\begin{table}
\centering
\begin{tabular}{|c|c|c|}
\hline
$\Delta x / \lambda_{D_e}$ & $\Delta x / a_{ee}^w$ & $N_c$ \\ \hline

0.3 & 0.088 & 0.00016 \\ \hline
0.5 & 0.147 & 0.00072 \\ \hline
0.9 & 0.264 & 0.0042 \\ \hline 
1.5 & 0.440 & 0.0194 \\ \hline
3.0 & 0.881 & 0.1554 \\ \hline
5.0 & 1.467 & 0.719 \\ \hline
\end{tabular}
\caption{Grid spacing and average number of macroparticles per cell for the numerical results shown in figure \ref{fig:ACH_Te_vs_t}.}
\label{tab:table_ACH}
\end{table}

\begin{figure}
    \includegraphics[width=8cm]{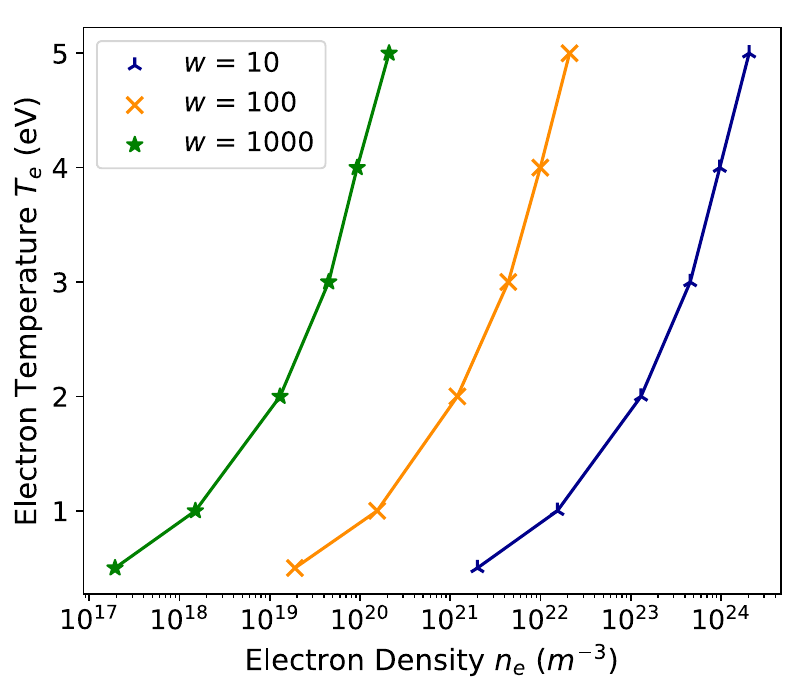}
    \caption{Artificial correlation heating limit curves for different macroparticle weights. The allowed operating region for PIC simulations is located on the left side of each curve.}
    \label{fig:Te_ne_different_w}
\end{figure}

To gain a more nuanced understanding of artificial correlation heating, the following methodology was adopted. First, the grid resolution was set to $0.5\lambda_{De}$ across all simulations. A macroparticle weight was then fixed at a specific value. An initial electron temperature was selected, and different simulations were run varying the electron density until the change in $T_e$; $\Delta T_e = T_e^{\max} - T_e(0)$; due to ACH amounted to 5\% of the initial electron temperature. Here, $T_e^{\max}$ corresponds to the electron temperature at $1.5\omega^{-1}_{pe}$. The corresponding value of $n_e$ was recorded, and the process was repeated for different initial electron temperatures. Each data point corresponding to the set of parameters $(n_e,T_e,w)$ represents an average across 50 simulations with different initial random seeds. This process was repeated for different macroparticle weights. Therefore, for each $w$, there exists a curve in the $(n_e,T_e)$ parameter space that demarcates the limit of the applicability of PIC to restrict ACH to no more than 5\% of the initial electron temperature. The region in which PIC is accurate lies to the left of the corresponding curve. Figure \ref{fig:Te_ne_different_w} shows the limiting curves for macroparticle weights of 10, 100 and 1000. As it is shown, the larger the macroparticle weight, the more limited is the density regime that can be simulated using PIC at a given temperature. This effect becomes more pronounced at smaller temperatures.

\begin{figure}
    \includegraphics[width=8.5cm]{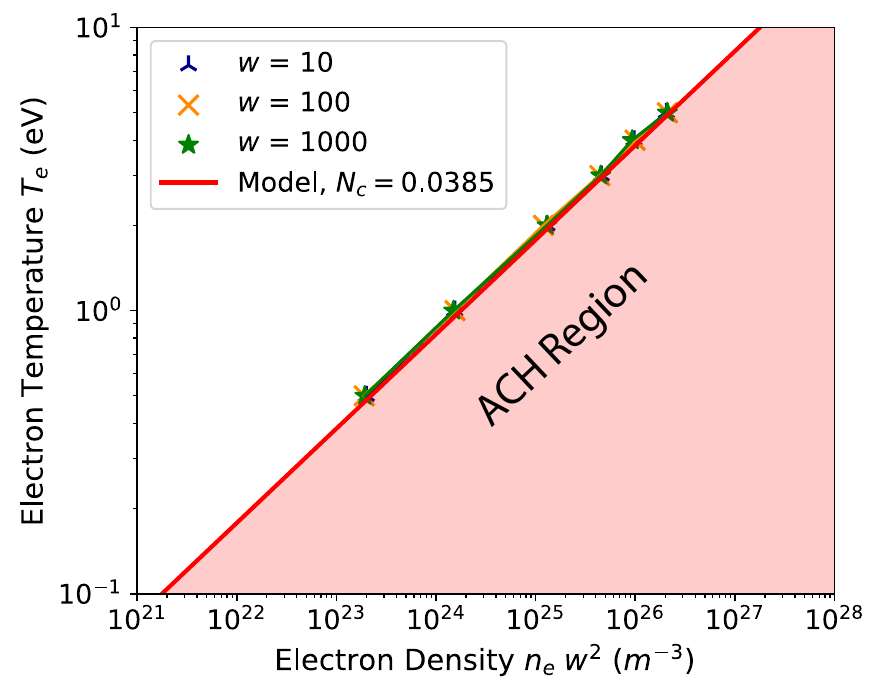}
    \caption{Limiting curve for ACH region from the model described in equation (\ref{eq:Te_vs_new2}) and numerical results from 3D-PIC simulations.}
    \label{fig:ACH_model}
\end{figure}

For the results shown in figure \ref{fig:Te_ne_different_w}, the input variables for the PIC simulations were $(n_e,T_e,w,N_x)$ where $N_x$ is the number of nodes in each direction and sets the size of the domain $L = (N_x - 1) \Delta x$. The grid spacing $\Delta x$ was chosen as $\approx f \lambda_{D_e}$ where $\lambda_{D_e}$ is the electron Debye length at the corresponding electron density and temperature and $f\approx0.5$. The number of macroparticles in the domain is given by $N_e = n_e\;L^3 / w$. The average number of macroparticles per cell is then $N_c = N_e / N_x^3$. Replacing the previous expressions in $N_c$, and assuming that $N_x
\gg 1$,
\begin{equation}
    N_c = \frac{f^3}{w} n_e \left ( \frac{\epsilon_0 k_B T_e}{e^2 n_e} \right )^{3/2}.
    \label{eq:Nc}
\end{equation} Solving equation (\ref{eq:Nc}) for $T_e$,
\begin{equation}
    T_e = \frac{N_c^{2/3}}{f^2} \left ( \frac{e^2}{\epsilon_0 k_B} \right ) (n_e w^2)^{1/3},
    \label{eq:Te_vs_new2}
\end{equation} gives an expression for the minimum electron temperature that can be simulated given the parameters $(N_c,\Delta x / \lambda_{D_e},n_e,w)$. This minimum electron temperature directly depends on the desired number of macroparticles per cell $N_c$ and the factor $n_e w^2$. It is observed that the chosen condition of a 5\% temperature increase due to ACH, used as a criterion in the data shown in figure \ref{fig:Te_ne_different_w}, is equivalent to a number of macroparticles per cell $N_c \approx 0.0385$ for all simulations. When this condition is applied to equation~(\ref{eq:Te_vs_new2}) with $f=0.5$, the numerical data align with the curve defined by equation~(\ref{eq:Te_vs_new2}) in the $(T_e,n_e w^2)$ parameter space, as illustrated in figure \ref{fig:ACH_model}.

On the right side of the limiting curve depicted in figure \ref{fig:ACH_model}, artificial correlation heating significantly influences the  electron temperature within a timescale dictated by the electron plasma period. In addition, on a longer timescale, PIC heating arises even when the Debye length is resolved due to the use of a limited number of macroparticles per cell—a direct consequence of the specific density, temperature regimes, and macroparticle weights employed. This region is not advisable for simulations, as even partial adherence to these conditions within the simulation domain could dramatically alter local electron dynamics. These findings considerably constrain the applicability of PIC simulations in scenarios of moderate electron temperatures and high ionization fractions or electron densities$-$an effect amplified by the high macroparticle weights inherent to the $n_e w^2$ scaling shown in equation~(\ref{eq:Te_vs_new2}). In addition, it is important to remark that ACH is not a consequence of non-energy conserving PIC schemes. ACH occurs as a consequence of the artificial increase in the Coulomb potential energy due to a large macroparticle weight. Hence, changing from a momentum-conserving to energy-conserving PIC scheme will not change the observed ACH at a given grid resolution. This is contrary to other sources of numerical heating, where for example, changing the PIC scheme can affect the growth rate of PIC heating.

It is vital to emphasize that in simulations where the conditions conducive to artificial correlation heating could potentially be met, ACH initiates a positive feedback loop that enhances its effect. Specifically, once ACH occurs in simulations where ionization of a neutral gas is included, the resulting numerical increase in electron temperature can trigger nonphysical ionization events, which subsequently elevate the electron density. This in turn provokes further ACH, pushing the simulation conditions towards the right side of the diagram depicted in figure \ref{fig:ACH_model}. The only way to evade ACH is by maintaining multiple macroparticles per cell. However, given certain conditions such as $(n_e , T_e)$, a fraction of the Debye length to resolve, and a macroparticle weight, these parameters immediately determine the number of macroparticles per cell. Consequently, the ACH curve depicted in figure \ref{fig:ACH_model} represents the limit of applicability of PIC simulations for electrons. This holds true even when the physical coupling strength corresponds to the weakly coupled regime. The results shown here demonstrate that PIC simulations of typical plasmas (e.g. capacitively coupled plasmas conditions) end up requiring lower macroparticle weights than one might expect in order to maintain ``many'' computational particles per element if the element is sized to be roughly the Debye length.

\subsection{ACH temperature increase}\label{subsec:ACH_T_increase}

The temperature increase due to ACH is similar to DIH but induced by an artificially enhanced coupling strength. The increase in electron temperature, accompanied by a decrease in the effective coupling strength, matches what was previously observed in MD simulations of DIH in \cite{Acciarri2022}. Thus, the DIH model used to predict the maximum temperature shown in equation~(\ref{eq:Ti_max}) can be used here if the grid spacing is small enough to well resolve the inter-macroparticle spacing. Replacing the charge and average interparticle spacing by its macroparticle counterparts and including the scaling of the macroparticle temperature,
\begin{equation}
    T_e^{\max} = \frac{1}{1.5} \frac{e^2}{4\pi\epsilon_0k_B} \frac{1}{a_{ee}} w^{2/3},
    \label{eq:ACH_T_max}
\end{equation} where $e$ and $a_{ee}$ are the physical charge and average interparticle spacing. Here, the numerical factor 1.5 is obtained from an average of PIC simulations where ACH is observed and corresponds to the minimum value of the effective coupling strength. Using the values for the simulation shown in section \ref{subsec:ACH_gamma}, the predicted maximum temperature is approximately $\approx$ 7.74 eV, close to the observed value of 7.65 eV.  It is important to underscore that equation~(\ref{eq:ACH_T_max}) is useful only when the macroparticle coupling strength is larger than $\approx 1.5$. For $0.2<\Gamma_{ee}^w<1.5$, ACH would still happen and influence the temperature observed in PIC, but to a lesser degree. In order to predict the change in temperature in the later case, a model based on conservation of energy should be used and will be developed in future work.

\section{Conclusions}

Particle-In-Cell (PIC) simulations have been a long-standing, essential tool in plasma modeling due to their ability to make kinetic simulations accessible at device-relevant scales. Currently, there is a concerted effort to extend the application of PIC simulations to model plasmas at atmospheric pressure. However, it has been shown that ions exhibit strong correlations at atmospheric pressure conditions \cite{Acciarri2022}. Here, we show that standard PIC simulations are only applicable to weakly coupled plasmas. Furthermore, when used to model strongly coupled plasmas, such as ions at atmospheric pressure, PIC requires additional constraints to capture phenomena like disorder induced heating. Firstly, a fraction of $0.1$ of the average interparticle spacing needs to be resolved in order to observe DIH, which on average results in less than one macroparticle per cell. Secondly, a unity macroparticle weight must be used. If the macroparticle weight is larger than the unity, it significantly influences the Coulomb potential energy due to a numerical localization of the electric charge, enhancing the electric field and the effective coupling strength, thus influencing the ion dynamics and temperature. These constraints considerably increase the computational cost of PIC for strongly coupled plasmas, rendering it impractical for device scale modeling. 

In addition, if PIC is operated with these additional constraints, PIC heating is inevitable due to the presence of less than one macroparticle per cell and high coupling strength. Attempts to reduce the growth rate of PIC heating with filtering methods or high order shape functions delocalize the charge density, artificially decreasing the coupling strength of the system and influencing the ion dynamics and temperature. Moreover, it is found that PIC simulations with reduced spatial dimensions are not suitable for strongly coupled plasmas. The scaling of the physical density in 1D or 2D simulations artificially decreases the average interparticle spacing which needs to be resolved to capture the strong correlations. A primary conclusion drawn from the presented results is that standard PIC simulations, are not suitable for strongly coupled plasmas. 
This places constraints on their applicability to atmospheric pressure plasmas to a sufficiently low density regime. 

Furthermore, while electrons are weakly coupled at atmospheric pressure, it is shown that the macroparticle coupling strength is influenced by the weight used in PIC simulations. A combination of large macroparticle weight and electron density can lead to a macroparticle coupling parameter larger than unity which induces artificial correlation heating. If the cell size is increased under these conditions, ACH can be reduced. A model for ACH was described that shows a good agreement with numerical results from PIC simulations and a general diagram was shown in the $(T_e,n_e w^2)$ parameter space to show the limit of applicability of PIC simulations.

Finally, we mention that although only pulsed discharges were considered here, DIH (or ACH) is also present in a continuously operating steady-state scenario. Consider a discharge that is lit by an initial ionization pulse, then subsequently obtains a steady-state in the presence of losses due to a continuous ionization source. 
The power released due to DIH at the beginning will be large because of the introduction of a large population of ions in an initially uncorrelated state. This will lead to the large initial increase in the temperature over a short time-period characterized by the ion plasma period, as described above. However, the ions created at the slower ionization rate of the steady-state discharge will also heat due to DIH because they are also introduced from a random distribution. 
Since they are introduced at a slower rate, the associated heating power will be less. But energy input per ion created from a random distribution will be the same, and therefore DIH will still set the steady-state temperature in the presence of an ionization source and sink. Thus, DIH is expected to be just as relevant to steady-state discharges. We also mention that in the case of ACH, if electron impact ionization collisions are included, this can induce a numerical instability that artificially increases the electron density exponentially over time. This complex interplay of ACH and ionization will be detailed in an future publication~\cite{Acciarri2024_ACH}.

\section{Acknowledgements}

The authors thank Dr.~Matthew Hopkins for helpful discussions. 
This work was supported in part by the US Department of Energy under award no.~DE-SC0022201, and in part by Sandia National Laboratories. 
Sandia National Laboratories is a multi mission laboratory managed and operated by National Technology and Engineering Solutions of Sandia, LLC., a wholly owned subsidiary of Honeywell International, Inc., for the U.S. Department of Energy’s National Nuclear Security Administration under Contract No. DE-NA0003525. This article describes objective technical results and analysis. Any subjective views or opinions that might be expressed in the paper do not necessarily represent the views of the U.S. Department of Energy or the United States Government.

\appendix

\section{Interpolation functions in PIC simulations}\label{sec:appendix}

The shape functions of order 4 and 6, used for the analysis on PIC heating detailed in section \ref{subsec:DIH_PIC_heating_interpolation}, are provided below.\cite{ABE_1986_shape_functions}
The order-4 shape function is
\begin{equation}
  W^{(4)}(x) = 
    \begin{cases}
      \frac{2}{3} - |x|^2 + \frac{|x|^3}{2}& \text{for } 0 \leq |x| \leq 1, \\
      \frac{1}{6}(2-|x|)^3 & \text{for } 1 \leq |x| \leq 2, \\
      0 & \text{otherwise },
    \end{cases}
    \label{eq:W_4}
\end{equation}
and the order-6 shape function is
\begin{flalign}
  & W^{(6)}(x) =  \nonumber \\ 
  & \begin{cases}
        \frac{1}{60}(33 - 30|x|^2 + 15|x|^4 - 5|x|^5) & \text{ for } 0 \leq x \leq 1, \\ \\
        \begin{aligned}
          &\frac{1}{120}(51 + 75|x| - 210|x|^2 + \\ &\qquad 150|x|^3  - 45|x|^4 + 5|x|^5 )
        \end{aligned}
        & \text{ for } 1 \leq x \leq 2, \\ \\
        \frac{1}{120}(3-|x|)^5 & \text{ for } 2 \leq |x| \leq 3, \\ \\
        0 & \text{ otherwise}.
      \end{cases}
  \label{eq:W_6}
\end{flalign}

\bibliography{references.bib}

\end{document}